\documentclass[acmsmall]{acmart}
\AtBeginDocument{%
  }

\setcopyright{cc}
\setcctype{by}
\acmJournal{PACMHCI}
\acmYear{2026} \acmVolume{10} \acmNumber{1} \acmArticle{GROUP16}\acmMonth{3} \acmDOI{10.1145/3799451}

\begin{document}

\title[The Sense of Misinformation Can Harm Local Community]{The Sense of Misinformation Can Harm Local Community: \\A Case Study of Community Conflict}

\author{Jiyoon Kim}
\affiliation{%
\institution{Pennsylvania State University, University Park, PA}
  \city{University Park, PA}
  \country{United States}}
\email{jxk6167@psu.edu}
\orcid{0009-0006-0627-5373}

\author{Jie Cai}
\affiliation{%
\institution{Tsinghua University, Beijing}
  \city{Beijing}
  \country{China}}
\email{jie-cai@mail.tsinghua.edu.cn}
\orcid{0000-0002-0582-555X}

\author{Srishti Gupta}
\affiliation{%
\institution{University at Albany, SUNY}
  \city{}
  \country{United States}}
\email{sgupta4@albany.edu}
\orcid{0000-0002-2354-0507}

\author{John M. Carroll}
\affiliation{%
\institution{Pennsylvania State University, University Park, PA}
  \city{University Park, PA}
  \country{United States}}
\email{jmc56@psu.edu}
\orcid{0000-0001-5189-337X}

\renewcommand{\shortauthors}{Jiyoon Kim, Jie Cai, Srishti Gupta, and John M. Carroll}

\begin{abstract}
During community decision-making and civic collaboration, conflicts can escalate when people suspect misinformation. We introduce the concept of \textit{sense of misinformation} as experiencing someone's language or behavior as misinformation when it is not, that is to say when no falsehood is involved. Misinformation and sense of misinformation feel similar and can have similar social consequences; but sense of misinformation rests upon a mistaken perception of someone else's information as false. Through a case study of a casino proposal in local community, we examine how sense of misinformation developed over time during a contentious civic process through key factors (i.e., miscoordination governance, miscommunication between local government and citizens, and conflict and the breakdown of civic discourse), undermining trust and community democracy. Distinguishing between misinformation and sense of misinformation presents a challenge, but it is important. We contribute a conceptual distinction to the misinformation literature by identifying this distinct phenomenon and discuss ways to help communities recognize and repair such misattributions. Finally, we discuss design approaches for mitigating sense of misinformation.
\end{abstract}



\received{July 2025}
\received[revised]{November 2025}
\received[accepted]{December 2025}

\maketitle

\section{Introduction}

Community conflict has long been an important topic in human-computer interaction (HCI). It plays a constructive role in strengthening community bonds by helping define shared identities, collective pursuits, and common ideals through the presence of contrasting ideologies or information \cite{disalvo2009design}. Instead of stifling or ignoring, conflict can foster the expression and negotiation of opposing views and beliefs, helping communities grow and evolve \cite{disalvo_community_2011}. Yet, conflict can also create polarization and weaken community bonds and trust \cite{coleman_community_1957}. In HCI, community conflict can motivate future directions for action and design trajectories that support community development \cite{disalvo_community_2011}.

Community conflicts arise from various issues such as land development \cite{de_jong_understanding_2021} and political elections \cite{hill_collins_new_2010}, both of which can generate polarized opinions and contentious debate. Casino development, in particular, has long been a controversial issue in local communities, creating tension between local governments and residents \cite{lee_residents_2010} and prompting debate over its potential economic benefits and harmful consequences for local communities \cite{rephann_casino_1997}, local businesses \cite{seelig_place_1998}, and organizations affected by the development project \cite{seelig_school_1998}. However, this issue has received little attention in HCI. In this case study, we examine community conflict surrounding a casino development project in a small town.

Local communities are at the fundamental level of social organization, encompassing towns and municipalities where people live. Unlike larger entities such as cities or countries, local communities often enable closer connections and more intimate relationships among residents. In these settings, neighbors can engage with one another directly, creating a distinctive social environment \cite{carroll_wild_2013, kwon_localized_2020}. Casino development is a controversial issue that can intensify community conflict and generate misinformation \cite{barrow2008persistence}. For communities confronting a casino proposal for the first time, it is not a routine policy disagreement but an unfamiliar event that requires residents to navigate significant uncertainty. To understand how a local community manages information practices surrounding a contentious local issue, we interviewed local government leaders and citizens about their information practice. We address the following two research questions:

\begin{itemize}
       \item How do local community stakeholders form and build their arguments based on information?
       \item What challenges arise in information practices among stakeholders regarding the casino project in the local community?
\end{itemize}

In this case study, we found that community stakeholders mistakenly perceive others' information as misinformation when there is no actual misinformation (i.e., false information \cite{cook2015misinformation}). We introduce this phenomenon as \textit{sense of misinformation} (SoM). In this work, sense of misinformation developed over time through multiple weaknesses in coordination and communication between the state government, local government, and the citizens (e.g., citizens opposing casino development created their own social media group to disseminate information and support collective action). Distinguishing SoM from misinformation is important for the HCI community because, even when no falsehood is involved, SoM can undermine community trust, cooperation, and dialogue. In our case study, these misattributions of misinformation contributed to fractured communication, weakened community bonds, and further confusion about identifying misinformation. This work contributes to the GROUP and HCI communities in three ways. First, we identify \textit{sense of misinformation} as a distinct phenomenon from misinformation per se. Second, we suggest that SoM requires different forms of repair from those used to address misinformation. Finally, we discuss design approaches that may help mitigate sense of misinformation.


\section{Related Work}

\subsection{Casino Development and Community Conflict}

Since the 1980s, casino development in the United States has expanded rapidly as a strategy for economic development. However, its introduction into local communities has been highly controversial \cite{rephann_casino_1997}, often provoking intense conflict because of sharply divergent views about its potential benefits and harms. From the government's perspective, casino development is often framed as an economic opportunity that can create jobs, increase tax revenue, and attract tourists \cite{national1999pathological}. Conversely, some scholars have argued that these benefits are substantially overstated and that harms such as gambling addiction are more significant than is generally recognized \cite{cosgrave_gambling_2001, lee_residents_2010}.

Casino development has also generated significant public concern in many communities \cite{long_early_1996, pizam_perceived_1985}. Citizens have often expressed strong concerns about potential increases in gambling addiction, local crime (e.g., theft), and changes to the community's image \cite{balzarini_gentrification_2016, calvano_hitting_2010, nichols_impact_2013}. Residents have also expressed concern about local economic consequences, particularly the possibility that casinos may draw customers away from nearby businesses \cite{seelig_place_1998}. These negative associations surrounding casino development can provoke substantial community resistance and spark severe controversy.

\citet{coleman_community_1957} argued that conflict can weaken community bonds. Casino development is a relatively new phenomenon in many U.S. communities \cite{akee2015indian, o2022impact, nichols2002community}, including the community in which this research was conducted. Unlike more routine policy disagreements, such as school budgeting \cite{tracy2007discourse}, casino development is often an unfamiliar issue that requires community members to navigate uncertainty while also confronting deep value conflict, for example, between the promise of economic development and expanded leisure opportunities, on the one hand, and concerns about social harm, addiction, and morality, on the other \cite{national1999pathological}. While prior research has examined community conflict in various contexts (e.g., \cite{gupta2021not, clegg2024navigating}), relatively little work has explored such new and value-laden conflicts in relation to local casino-related information practices. 

\subsection{Local Government and Community Information Practice in HCI}
HCI scholars have developed technologies to support civic engagement and democratic practice across multiple levels of government in the United States, including systems for collecting public input at the federal level \cite{faridani2010opinion}, participatory platforms at the state level \cite{kriplean_supporting_2012}, and civic technologies for local and municipal governments \cite{jasim2021communityclick}. Together, this body of work shows how digital systems can make government more accountable and participatory, while highlighting the role of sociotechnical systems in shaping civic infrastructure \cite{dantec2013infrastructuring, gordon_augmented_2011}.

Local governments face distinctive challenges because officials must directly mediate interactions between citizens and government. Providing essential information to the public is critical to the success of local projects, and miscommunication can contribute to project failure \cite{corbett_problem_2018, hong_information_2018}. In addition, local governments' internal operations and coordination with other levels of government are crucial for balancing limited resources with urgent public demands, especially during crises \cite{meyerhoff_nielsen_digital_2023, peters_challenge_2018}. To address these challenges, municipalities increasingly adopt information and communication technologies to improve transparency and responsiveness in their communication with citizens \cite{lorenzi_utilizing_2014}. Ultimately, however, effective local governance depends on officials' ability to create opportunities for public engagement and dialogue during policymaking while upholding core democratic values \cite{couto_empowered_2005, overney2025boundarease}. In practice, municipalities often struggle to gather meaningful public input \cite{corbett_problem_2018}.

HCI scholars have also examined a wide range of civic technologies intended to support civic engagement \cite{kavanaugh2005information, fan2022access}. Civic engagement refers to community members' voluntary actions, undertaken individually or collectively, to participate in decision-making and address public issues, and it is widely recognized as a foundation of democratic practice \cite{wiegard2025cultivating}. Meaningful engagement requires citizens not only to be aware of ongoing events \cite{baker_critical_2005}, but also to understand local policy issues such as budgeting \cite{kim2015factful} and urban planning \cite{lourencco2007incorporating}. Yet technologies designed to increase participation often struggle to sustain civic engagement over time \cite{maas2023hubbel}. These limitations arise from challenges such as unequal access to information \cite{crivellaro_infrastructuring_2019}, participation fatigue, and the reproduction of existing power structures \cite{dantec_infrastructuring_2013}. Moreover, civic technologies can create transactional forms of democracy in which participation is reduced to one-off interactions rather than ongoing dialogue. This suggests that the value of civic technology lies not simply in producing more tools, but in supporting the social infrastructure of democracy. Trust must therefore be cultivated to enable meaningful and participatory democratic practice \cite{mccord_beyond_2023, corbett_problem_2018}. To support this, civic systems should enable two-way information flows between governments and communities so that citizens' perspectives are meaningfully incorporated into decision-making \cite{overney2025boundarease}.

Although prior research has examined relationships between local governments and the public \cite{corbett_problem_2018, hong_information_2018}, as well as technologies that support local government management \cite{lorenzi_utilizing_2014} and decision-making \cite{couto_empowered_2005, overney2025boundarease}, relatively little attention has been paid to how information practices shape community dynamics during severe conflict. 
Whereas much prior work has focused on improving the collection of public input \cite{corbett_problem_2018}, our case study suggests a different problem: how information practices can intensify tensions within an already highly engaged public. More broadly, although civic technology research reaches many contexts, our case study is grounded in the specific setting of U.S. local governance (i.e., municipal- and township-level government), which operates within distinct state-level bureaucratic structures.

\subsection{Misinformation in Local Communities} 

Misinformation has been extensively studied in both online and offline contexts. Research on online misinformation has largely focused on large-scale public social media platforms across a range of domains, including crisis management \cite{huang2015connected}, public health \cite{wang_understanding_2023}, and political conflict \cite{mazepus2023information}. For example, during the 2016 U.S. presidential election, the Russian government supported coordinated propaganda efforts on Facebook and Twitter \cite{persily2020social,shu2020mining, pierri2023propaganda}. Beyond public social media, HCI research has also examined the distinct challenges of misinformation in private messaging environments such as WhatsApp \cite{pasquetto2022social} and LINE \cite{lee2025countering}. To mitigate misinformation, scholars have proposed a variety of interventions, including fact-checking tools \cite{gamage2022designing}, personalized nudging strategies \cite{biselli2025mitigating}, and chatbots designed to serve as neutral and productive moderators \cite{lee2025countering}.

Another line of research has extended beyond online environments to examine how misinformation shapes individual and collective behavior offline. Studies have shown that misinformation can fuel social polarization and mobilize collective action \cite{hasan2025role}. Misinformation can also contribute to broader social movements and ideological divides, as seen in the rise of the anti-vaccine movement, which has raised concerns about future global health crises \cite{aghajari2023s}. At the individual level, misinformation often becomes intertwined with preexisting values and belief systems, contributing to the adoption of conspiracy theories such as flat-earth beliefs \cite{engel2023learning}. These behavioral effects are closely tied to underlying cognitive processes. Confirmation bias, the tendency to seek information that aligns with one's existing beliefs \cite{lewandowsky2017beyond}, and cognitive dissonance, the discomfort that arises when encountering conflicting information \cite{taber2006motivated}, can both make individuals more susceptible to misinformation and more resistant to corrective evidence. These dynamics also extend to community settings. Confirmation bias can intensify community divisions by reducing openness to differing viewpoints \cite{lewandowsky2017beyond, schlosser2021confirmation}. To reduce the discomfort of disagreement, individuals may dismiss opposing perspectives and rationalize their own positions, thereby reinforcing divisions and deepening conflict \cite{acharya2018explaining}.

While prior research has explored the impact of misinformation on democracy and social division \cite{drolsbach2023diffusion, reglitz2022fake}, relatively little attention has been paid to how misinformation becomes embedded in social contexts, particularly in local communities. Local communities differ from both global and personal contexts in that they are characterized by close neighborhood ties, familiarity among residents \cite{carroll_wild_2013, kwon_localized_2020}, and reliance on local social media platforms for information practices \cite{masden2014tensions, lopez2015lend}. When a controversial issue such as casino development emerges, misinformation can add another layer of complexity \cite{barrow2008persistence}. In this case study, we examined misinformation in the local community context of a casino development project to better understand community information practices. Notably, we observed cases in which perceptions of misinformation persisted even in the absence of actual misinformation, a phenomenon we call \textit{sense of misinformation}.

\section{Method}

A case study allows researchers to identify and contribute to knowledge about individual, group, organizational, and broader social phenomena. By examining contemporary events, researchers can use case studies to address \textit{how} and \textit{why} questions about social phenomena \cite{yin_case_2009}. To understand community information practices, we conducted a case study of stakeholder information practices surrounding a contemporary casino development proposal within a local community.

The research site is a university town that serves as a regional economic center, making it a promising candidate for casino development. We selected this community for two primary reasons. First, its status as a viable site for casino development had generated significant community conflict, providing a highly relevant context for studying stakeholder information practices. Second, the researchers were members of the community (i.e., local residents), which offered an important methodological advantage. This insider position helped foster trust and cooperation, both of which were necessary for our qualitative approach and enabled more in-depth interviews with community stakeholders.

The research team consisted of multiple researchers, all of whom had lived in the community for periods ranging from 2 to 20 years. Two researchers attended a public hearing in September 2022 to understand the current status of the casino project. Following this visit, we began recruiting participants. All participants volunteered to take part in the study. Semi-structured interviews were conducted via Zoom and audio-recorded between October 2022 and February 2023. Interviews lasted between 29.5 and 84.07 minutes ($M = 49.4$). This study was reviewed and approved as exempt human subjects research by university's Institutional Review Board (IRB).

\subsection{Interview Study Design \& Participant Recruitment}

We recruited 10 citizens and 4 local government officials using three approaches: email recruitment, local social media platforms, and snowball sampling \cite{naderifar_snowball_2017}. Citizens opposing the proposed casino were recruited through publicly available email addresses on the local government's website, which publishes emails sent by citizens to government officials regarding the proposed development. This publicly available information enabled researchers to directly contact citizens who opposed the casino. To recruit citizens who favored the casino, we reached out through Nextdoor, a platform that enables local communities to share information and support one another \cite{NextdoorAbout}.

We aimed to recruit participants representing both sides of the issue to ensure a more balanced perspective. The underrepresentation of pro-casino participants may reflect the fact that the township's official website already documented the pro-development position extensively, which may have reduced the perceived need for supporters to voice their views separately. We also contacted citizens who had posted or commented about casino development on local social media platforms through direct messages. Finally, we used snowball sampling \cite{naderifar_snowball_2017} by asking interviewees to introduce others who either opposed or favored the casino development. Citizens who opposed the casino were relatively easy to reach because many of their email addresses were publicly available on the government website.

One participant (P1) was recruited through the local government's website using a publicly available email address from a citizen opposing the proposed casino. This participant introduced researchers to P2, the unofficial leader of a citizen opposition group. P2 then connected with four additional participants who opposed the casino (P3, P4, P5, and P6). The remaining four participants (P7, P8, P9, and P10) were contacted directly through messages on a local Facebook group or on Nextdoor. We contacted all local government officials using publicly available email addresses listed on their official websites. We concluded recruitment after reaching thematic saturation, when no new themes or patterns emerged from participants' utterances \cite{fusch_are_2015, saunders_saturation_2018}.

For \textit{local government officials}, we asked about the origins of the casino development proposal, the local government's role and perspective in relation to it, and the practices and challenges involved in disseminating casino-related information to the community. We also asked about the types of information officials used for decision-making and whether they were aware of how community members consumed casino-related information. In addition, we explored how officials perceived the information currently used by the community. For \textit{citizens}, we asked where or whether they obtained casino-related information, whether they stayed up to date on the issue, and whether they compared such information with other sources or discussed it with others. We also asked about their views on the local government's information-sharing practices and their opinions of public hearings hosted by the local government. In addition, we examined their personal views on the casino development proposal, how they used information to form their positions, and how they shared information within the community.

\subsection{Data Analysis}

To analyze the interview data, we first imported all transcripts into Taguette\footnote{https://www.taguette.org} for collaborative coding. The analysis proceeded in two parallel ways. We conducted an independent iterative thematic analysis \cite{braun_thematic_2012} and affinity diagramming using a Miro board \footnote{https://miro.com/ap2p/} to perform organize themes. Initially, two researchers independently coded one transcript to familiarized themselves with data. 
The research team then met regularly to discuss the initial codes, clarify their definitions, and resolve disagreements through discussion until consensus was reached. Throughout the analysis, the team aligned on coding strategies, developed a preliminary codebook, iteratively refined codes, and collaboratively and  iteratively constructed categories and themes. Following refinement of the codebook, the first author conducted axial coding to further organize and relate the themes. All open codes were placed onto Miro, where participant utterances were visually grouped to identify emerging thematic structures. These groupings were reviewed by all researchers. Through multiple rounds of iterative thematic analysis \cite{braun_thematic_2012}, we identified overarching coherent themes.

\begin{table}[tp]
\begin{tabular}{llllll}
\hline
Participant ID & Representation & Positions & Participant ID & Representation      & Positions   \\ \hline
P1             & Citizen        & Opposed   & P9             & Citizen             & Support     \\
P2             & Citizen        & Opposed   & P10            & Citizen             & Neutral     \\
P3             & Citizen        & Opposed   & P11            & Township Official   & No position \\
P4             & Citizen        & Opposed   & P12            & Township Official   & No position \\
P5             & Citizen        & Opposed   & P13            & Township Official   & No position \\
P6             & Citizen        & Opposed   & P14            & Township Official   & No position \\
P7             & Citizen        & Support   & \\
P8             & Citizen        & Support   &                &                     &             \\ \hline
\end{tabular}%
\caption{Interviewee positions. Township official's explanation of no position: (P14) "Township did not take any action. We're not going to say, No, we don't want it; we're not going to say yes, we want it; we're just going not to take action (did not "opt-out").". }
\label{tab:my-table}
\end{table}

\section{Findings}

Throughout this paper, we use \textit{local government}, \textit{township}, and \textit{municipality} interchangeably.


\begin{table}[h!]
\centering
\caption{Estimated Timeline for the Casino Development Proposal}
\label{fig:casinotimeline}
\begin{tabular}{@{}p{1.5cm}p{2.5cm}p{8.5cm}@{}}
\toprule
\textbf{Phase} & \textbf{Date} & \textbf{Key Event} \\ \midrule
\textbf{Phase 1} & Mid-2000s & The State Gaming Board legalized casinos statewide. \\
 & 2016 & Township established zoning regulations allowing gaming establishments in commercial zoning districts. \\[3pt]

\textbf{Phase 2} & 2017 Oct & The state gave municipalities six weeks to ``opt out'' of casino development proposal. \\
 & 2017 (Oct--Dec) & Township held public hearing. \\
 & 2017 (Oct--Dec) & Township internal discussions regarding the casino development. \\
 & 2017 Dec & Township decided not to ``opt out.'' \\[3pt]

\textbf{Phase 3} & 2019 Jul & Township had a second opportunity to ``opt out'' during a 60-day period. \\
 & 2021 Aug & Community members began writting letters to the media opposing the casino. \\
 & 2021 Aug & The State Gaming Board held a public hearing. \\
 & 2022 Apr & Citizens opposing the casino launched online and offline petitions. \\
 & 2022 Jun & Citizens sent the petition results to the Gaming Board. \\
 & 2022 Jun & The Gaming Board closed the public comment period. \\
 & 2022 (Sep--Oct) & Multiple public hearings held to discuss aspects of proposed casino development. \\
 
\bottomrule
\end{tabular}
\end{table}

\subsection{RQ1: Process of Building Arguments about Casino Development Proposal}

\subsubsection{Citizens' Process of Building Their Argument}
\label{sec:citizens-argument}

As shown in \autoref{fig:casinotimeline}, the public remained unaware of the casino proposal until the initial public hearing in 2021. This lack of transparency fostered mistrust toward the developers and local government, leading some citizens to suspect that information had been intentionally withheld. P5 (against) described the developers as trying to \textit{``sneak it in under the radar.''} Similarly, P1 (against) said, \textit{``It seems like it's part of the process. Maybe... the developers and some of the council members don't want us to know. So they make it hard for us to find out. I kind of wonder that.''}

The public's lack of awareness of procedural processes heightened mistrust toward the township’s information-sharing practices. This situation provoked strong emotional responses and sparked opposing narratives amongst members ofthe public.

\paragraph{Public Collective Action and Collaboration}

In early 2022, the citizens began mobilizing collectively to oppose the proposed casino in the township. Our interviews suggested that opposition voices were more prominent than pro-casino voices.  P2 served as the unofficial leader of the opposition group and took charge of making a comprehensive website opposing the casino development. For example, P2 noted, 

\begin{quote}
\emph{As I did more and more research, I realized that there's a lot of information out there that many people in the community didn't know. I was finding it. I thought I might as well compile this and make it available so other people can easily locate the same information that I found. So it was probably early spring when I started putting that up. - P2 (against)}
\end{quote}

P2 was committed to posting one relevant news article everyday. To curate content, P2 reviewed recent local publications and newspapers, as well as reports from various news organizations, using ``casino development'' as a primary keyword. Although the site began modestly with just a few pages, it grew to 13 pages. P2's website primarily focused on the negative impacts of casino development, such as gambling addiction, potential crime, casino-related child neglect, human trafficking, and money laundering, citing cases from places like Las Vegas and Atlantic City. The primary sources consisted of online news articles that covered not only local community events but also incidents and activities at casinos in other states. Although the website was basic and lacked advanced technological features, its primary purpose was to effectively distribute information to the public. The website also received contributions from other citizens who corresponded with P2 and participated in various public efforts to oppose the casino. P3 explained how P2's research and website helped them understand the casino development process and form an opinion:

\begin{quote}
\emph{Yeah, as far as making me aware of what what some of the things were going on... [P2] did a lot of the research and so [P2] could provide that aspect of educating people. But [P2] was also so helpful in knowing how the process worked, how the approval process happened at the state with a gaming board and how the process worked with the local government. - P3 (against)}
\end{quote}

Once the public became aware of a casino plan, P2 and other citizens opposed to the development, employed several methods to raise awareness and gather support. They disseminated information through word of mouth and social media platforms such as Facebook, Nextdoor, and Twitter. This included posting updates and articles on their opposition website and on social media to call for action, reach a broader audience, and mobilize community members. They also notified the broader community about specific events, such as the township’s public hearings, to encourage other citizens to attend and provide input. The opposition group attended these hearings, either in person or via Zoom, to support and represent the citizens' interests.

The opposition group created both online and offline petitions to gather community support. The offline petition involved of door-to-door visits in surrounding neighborhoods to explain the issue and collect signatures. The online petition was posted on a popular local newspaper's website. Furthermore, the opposition group compiled an email list of petition participants and individuals from community churches who wanted to receive updates on the casino development. This email list allowed them to provide regular updates on the progress and activities of the opposition movement, as well as share announcements about relevant activities, such as attending township public hearings. Many recipients forwarded these emails to their friends and family, enabling the information to reach a broader network. As a result, the email list grew to approximately 1{,}100 people, demonstrating substantial reach within the community and supporting effective communication and mobilization against the casino development.

\paragraph{Perceptions of Bias in Other Stakeholders’ Information}
The opposition group perceived that the local government's information on its official website, including local and economic impact reports, was biased. P2 explained, \begin{quote}\textit{All this information... we're going to create so many hundreds of jobs, we're going to attract [billion] of revenue. They make it sound really good. But all of that has to come from somewhere... it's going to come from people in this community. - P2 (against)} 
\end{quote} 

According to some opposition participants, two surveys were conducted: one by a popular local news website and another by citizen opponents. A poll conducted by the news website asked visitors whether they favored or opposed the casino, and approximately 80\% opposed it. Opposition participants cited these results to emphasize strong community resistance, as P3 (against) explained: \emph{``It's very high... Even one of the local media sites did a poll... it was very high number like 85,90\%. were opposed to [casino development].''}


However, pro-casino citizens disputed opposition's claims and polls, arguing that opponents disseminated false information due to personal bias and that the opposition's arguments reflected personal concerns rather than valid arguments against casino development in this local area. For example, P7 noted,
\begin{quote} \emph{People who are opposed to it really don't know what the hell they're talking about. They act as if having a casino is going to cause all kinds of social problems. That’s just absurd (...) You're telling me this false information that 80\% oppose (...) The same people who oppose the casino are using the same arguments used for alcohol prohibition, prohibiting prostitution, and prohibiting marijuana. All of their arguments are flawed. - P7 (favor)} \end{quote} 

P7 labeled the 80\% poll as false information, not because they identified evidence of misinformation, but because they regard the opposition's moral values as fundamentally flawed and therefore perceived the survey poll as wrong. This is an example of \textit{sense of misinformation}.

Some pro-casino citizens noted that they often seek information in local newspapers, social circles, and on social media platforms like Nextdoor to understand the proposal's status, but they struggle to find unbiased sources. For example, P8 explained,

\begin{quote} \emph{It's hard to get actual information... It's really hard to find one place that is unbiased to get information just like anything today. Like real data, raw data that I can make my own decision about, it's not manipulated or biased towards, you know, an outcome that somebody's trying to make happen. - P8 (favor)}\end{quote}

Both groups also relied heavily on anecdotal evidence from personal social circles. P1 (against) said, \textit{``I have some anecdotal evidence that it's correct. In terms of that, pretty much everybody I've talked to, at least in my social circle, is against the casino.''} In contrast, P7 (favor) noted, \textit{``If you talk to anyone there. I don't know anyone personally who opposes it. Everyone I know is either in favor of it or they just don't care.'' } Overall, both those who favor and oppose the casino perceived the other side’s arguments as biased, misleading, and false.

\subsubsection{Local Government's Process of Building Their Argument}


Township officials used various types of information, such as local impact reports, economic impact reports, and personal observations gathered from field trips to other similarly sized community casinos, to formulate a compelling argument for why they concluded that the casino would have a positive impact on the local community. When the township decided not to opt-out, several factors affected decision-making, including preexisting zoning regulations. For example, one official (P12) said, \begin{quote}\emph{So 2017 was when the state said, ``opt out of casinos if you want''. For a number of reasons, we didn't. Largely just because we already had it in our zoning, a lot of the other municipalities didn't have it in [their] zoning. But we had a spot for it in our zoning already. }
\end{quote}

A consulting firm, hired by a gaming company to manage the day-to-day operations related to the proposed casino development, produced a local impact report and conducted economic and fiscal impact analyses. The local impact report used multi-step approach. First, the consulting firm analyzed the 2021 budget to understand how funds were allocated to municipal services. Then the firm spoke with local government administrators and managers about current services and how the casino could affect them. Based on this information, the firm performed a fiscal impact analysis to estimate the casino's overall impact on local government departments. This report aimed to understand how the casino would affect existing services, resources, and tourism. The economic impact report used input-output modeling to estimate the total activity impact of the proposed casino. The total economic impact was determined by combining direct expenditures, indirect economic activity and jobs, employee compensation, and induced economic activity and jobs.

After township officials reviewed reports, they concluded that the casino would have minimal negative impact on the local community. One official (P11) said, \emph{``I see minimal impact to community infrastructure wise, and there just isn't any data to support any incremental negative impact on a brick and mortar casino current in [state], under the current [state] casino regulations.''} Township officials also conducted a site visit to similarly sized casino in another community to evaluate its impact. P11 added, \begin{quote}\emph{We went down to [another town] and talked firsthand to everybody. Everybody told us the same thing - none of that stuff is happening there for many reasons. One, the casino has a lot at stake. If those things were happening there on that property, the casino would lose its license because of the very heavy oversight or regulation that [state] provides and requires for these. So it just wasn't happening.}\end{quote}

In 2017, officials internally reviewed and analyzed information provided by the state government to inform decisions about casino development. By 2021, they had conducted field visits to casinos in similarly sized communities and analyzed data on local impacts. These analyses, together with economic impact reports, were documented and published on the local government's website as factual resources.

\subsection{RQ2: Challenges in Stakeholder Coordination and Communication}
\subsubsection{Miscoordination Governance}
\label{find:miscoordination} 
\paragraph{State Government Provides Tight Deadline on Local Government for Early Stage Decision-Making}
As shown in \autoref{fig:casinotimeline}, in October 2017 the state gave municipalities six weeks to decide whether to opt out of casino development. Communication between state and local government relied largely on conventional channels. P14 noted, \textit{``Routine matters with each level of government, the primary means of communication remains via mail, phone, or email. We do not typically utilize other online methods for that type of communication.''}

During this period, most residents remained unaware of the proposal. Township officials therefore relied primarily on state-provided information about the potential benefits and risks of casino development and held internal discussions, ultimately deciding not to opt out. As one township (P12) official described,
\begin{quote}\emph{Back in 2017, the state gave municipalities six weeks to opt out of having a casino in their municipality, so a lot of the municipalities in the [county] opted out; we did not. We let that pass. We did have a public hearing about it; nobody came to speak.}\end{quote}

When a second 60-day opt-out opportunity occurred in July 2019, the township took no official action, thereby leaving casino land use permissible under existing policy. P12 noted that residents may not fully understand the roles of different levels of government or where decision-making authority resides. This suggests that intergovernmental processes can be largely invisible to constituents, limiting public understanding and increasing uncertainty. State-imposed deadlines may have further contributed by compressing decision-making timelines and reducing opportunities for information gathering and public input. As a result, some citizens interpreted their late awareness of the casino issue as intentional concealment by the local government (see \ref{sec:citizens-argument}).

\subsubsection{Miscommunication between Local Government and Citizens}
\label{find:miscommunication}
\paragraph{Local Government's Challenges of Communication Infrastructure}

During the six-week decision-making window, the township used multiple channels to disseminate information, including its website, social media platforms (e.g., Facebook and Twitter), and local newspapers, to publicize public hearings about opt-out decision. One official (P11) described the issue as widely covered at the time:
\begin{quote}
\emph{It was high-profile news because it was happening all over the state. All the municipalities in [county], going through that opt-out decision, and it was covered in the media; it was news in the local newspapers. Just like anything else that we debate, deliberate, and decide on, it was shared in a public format, like any of the other decisions we make.}
\end{quote}

As the scheduled meeting approached, township posted official agenda (including dates, times, and locations) roughly a week in advance. However, citizen involvement remained low during the six-week opt-out period, and public hearing attendance was sparse. Many residents attributed this low participation to limited awareness. Township officials also described challenges in providing timely, relevant information due to their crowded daily schedules and uncertainty about what residents would most care about. As one official (P11) noted, information sometimes ``slipped through the cracks'': \begin{quote}\emph{That's a challenge. How do you know what information people are interested in? And how do you push that out to them in a timely manner? People are just all very, very busy in their lives with other things. A lot of things slipped through the cracks.}\end{quote}

The township also relied on regular email updates that citizens had to actively subscribe to, and officials described difficulties maintaining subscriptions after a website transition (P12, township official). Meanwhile, some residents noted receiving no direct communication from the township and remaining unaware of the proposal (P6, against): 
\begin{quote} 
{\textit{We struggle with things like our website. When we changed website companies, they took all of our [email] subscribers, and we had to start over. Our email subscription is now super low because people aren't signing up again. I totally understand that; I was signed up, why do I have to sign up again? - P12}}
\end{quote} 

\begin{quote}
    \emph{I don't think I've ever in the 15 years... gotten an email, or any kind of contact from someone from the township saying, ``just so you know, this is happening, and we want you as a community member... to be aware of this''... Probably because they just think we don't care which most of the time we don't... But there have been some situations where we definitely do care. - P6 (against) }
\end{quote}

In addition to email updates, the township posted all casino-related information on its official website, including timelines, documents, public hearing recordings, public emails, and FAQs. Local government's crowded agendas, subscription barriers, and uneven outreach contributed to low awareness and limited participation, which in turn intensified community frustration and uncertainty about the proposal's trajectory.






\paragraph{Citizens Criticize Limited Transparency and Lack of Feedback from the Local Government}

After many citizens became aware of the casino development, divergent opinions made it challenging for the township to facilitate discussions with the community. To address public concerns and fears about the casino's potential impact, the township organized several public hearings in the fall of 2022. These hearings were intended to provide space for citizens to express their views and were held in a hybrid format, both in person and via Zoom. However, the local government struggled to balance open citizen expression with focused, relevant discussion, and citizens often received little feedback or left without understanding the government’s position after sharing their views.

Between August and October 2022, many citizens who opposed the casino proposal attended the township's public hearings and shared their opinions with township officials. Citizens acknowledged the local government's efforts to organize public hearings and provide opportunities for citizens to express concerns and perspectives. Still, discussions during these hearings were largely one-way, with limited feedback from township officials. Many citizens were dissatisfied with how township officials managed the public hearings and felt that officials did not adequately acknowledge the points they raised. P1 noted:
\begin{quote}
\textit{I felt empowered because there were so many people that had excellent points, so many members of the public. I didn't appreciate the attitude of the council members. They seemed hostile to me. They also didn't seem to be very interested in public comment. They seemed like they were going through the motions because they had to do it, not because they really cared what the public [say] -P1 (against).} 
\end{quote}

Citizens appreciated that officials organized emails from opposition groups on their website. However, officials did not provide any direct feedback or response. They also noted that township's lawyer had cautioned against openly communicating the level of opposition to the Gaming Control Board at the state level. Despite this legal advice, one participant believed that if the government had genuinely empathized with the community's strong opposition, officials might have been willing to take the risk of conveying those sentiments, even if doing so involved some legal risk. For example, P3 (against) said, \emph{``I don't feel they really understood concerns, or that they shared the concerns. I did understand that they were warned by their lawyer not to communicate...''}

A lack of responsiveness from officials, both during hearings and via email, fueled public frustration and created a sense that citizen concerns were being dismissed. This perceived lack of transparency led residents to seek information from opposition leaders and engage in collective action against the casino development. Ineffective communication further reinforced citizens' mistrust of the local government.

\paragraph{Municipality Could Not Act Due to Potential Litigation}
The township did not publicly express a clear position, maintaining that it could not take further action. As P3 observed during public hearings: \textit{``They didn’t express any opinion. They just kept saying there is nothing they could do.''}

Opposition residents urged the township to send a formal letter to the Gaming Control Board opposing the casino license. On October 6, 2022, township officials held a public vote via Zoom, during which four of the five officials voted against the motion, citing legal risk. The township's attorney warned officials that opposing the casino at this stage could trigger a potential \$20 million lawsuit from the casino developer. While citizens recognized the legal risk, many felt that the township failed to represent the community. For example, P5 noted,
\begin{quote}
 \emph{I don’t think the township leaders are adequately representing many people in the township, like me, who don't want this casino. I understand lawsuits are costly and used as a fear factor. They're just unwilling to change their mind out of fear, even though the people are clearly opposed. - P5 (against)}   
\end{quote}

\paragraph{Local Government's Concerns About Irrelevance and Misinformation in Citizens' Arguments}

Township officials described active community engagement during public hearings as valuable, but they also expressed concerns about the relevance and applicability of some citizen arguments. Officials maintained that their primary role was to gather facts and noted that they sometimes had to moderate hearings to ensure that discussions remained professional and focused on the proposal. As one official (P11) noted,
\begin{quote}
\textit{Our job is to gather facts for the most part. So I think you'll let some of that come out in the public hearing. But sometimes you have to stop people. Because they've now turned the public hearing into maybe a personal attack, or they're throwing out completely irrelevant data, or [personal] experiences. } 
\end{quote}

Many citizens raised concerns about potential negative impacts on the community, such as gambling addiction and increased crime, often drawing on examples from casino cities such as Atlantic City, Reno, and Las Vegas. However, township officials perceived that these comparisons were not applicable to the local case. They argued that much of the community's opposition was inapplicable in this context, as this state’s regulatory framework differs substantially from those of other jurisdictions. As one official (P11) noted,

\begin{quote}\emph{Data from 15, 20 years ago about gambling or crime data, based on an experience in Atlantic City. What we are doing in our exercise is looking and saying, is that really relevant to our situation? It's not relevant or applicable.}\end{quote}

A township official also raised concerns about misinformation within the community, which was often perceived as problematic or unreliable. One official (P12) described the challenge of informing new citizens who joined meetings without first reviewing the information already provided by the local government, requiring officials to repeatedly explain the same details to each new group of attendees. This task was made more difficult by the fact that these groups often came to meetings with preconceived notions.
\begin{quote}\emph{I do believe that they had a lot of good information. Unfortunately, as usual, it's mixed with misinformation, right? There's always another new group of people that you have to catch up with. And this is a complicated topic; it's hard for people to understand that we just don't have an on or off switch. }\end{quote}

A township official further noted that emotion could contribute to the creation of misinformation among citizens. When emotions run high, citizens may generate misleading claims to justify their fears, such as concerns about potential crime, gambling, or alcohol addiction. The council described some of what circulated as lacking factual evidence, as P12 noted,

\begin{quote} \emph{I do think that fear is its own misinformation creator. If there's something scary, people like to come up with all kinds of nonsense to support their fear. I'm not saying that everything that these people said was nonsense. A lot of it was very good, factual data, but some of it was absurd. Some of it was wacky; it just was not truthful.} \end{quote}

This suggests that the township official dismissed citizen input, perceiving their information as misinformation. This perception emerged when those arguments were rooted in moral concerns and differing community values, suggesting the presence of \textit{sense of misinformation} rather than false information. In a follow-up question, P12 was asked to provide examples of misinformation shared by community members. However, P12 could not recall a specific case and ultimately retracted the term, clarifying, \emph{``Um, I just think maybe it’s not misinformation...''} Although the official acknowledged that the term \textit{misinformation} had been a mischaracterization, they still felt that something was wrong with the information shared by citizens.

Another township official noted that this case reflects a broader trend: when governments provide factual information, they are often met with public skepticism and mistrust. This lack of confidence leads the public to favor subjective opinions over official data. Ultimately, the central challenge is one of credibility. Officials felt that this issue extends beyond local governance, reflecting a broader erosion of trust across all levels of government. For example, one official P14 said,
\begin{quote}\textit{
This is the challenge of local government, governments across the board, honestly. We've tried to share what we believe is all factual information and outline the process and potential development. The question becomes: do those receiving that information believe it? Everyone's entitled to their thoughts and opinions.}\end{quote}



\subsubsection{Conflict and the Breakdown of Civic Discourse}
\label{find:conflictbreakdown} 



\paragraph{How Offline Petitioning Became an Unintentional Threat}

Citizens opposing the casino took collective action through online and offline petitions. Opposition leaders and residents, motivated to protect and improve their community, went door to door to collect signatures, sometimes visiting multiple times. Although this approach was intended to build support, it unintentionally intimidated some pro-casino residents (P8), leading to discomfort and anxiety about further interaction: \emph{``I said, no, I want the casino. And asked why you don't want the casino? They couldn't tell me why except that it's bad.''}

\begin{quote}\emph{I'm seriously afraid of [them]. When people disagree with [them], they don't think your opinions are bad. They think you're a bad person instead of just your opinion. That's the way people think anymore. And I just don't want to be around people. If they're so passionate [and] they have to go door to door to try to block a casino. I don't want to face them. - P8 (favor)}\end{quote}

This perception led some casino supporters to feel that expressing their views marked them as bad people, rather than as individuals with legitimate differences of opinion. This atmosphere was further amplified by local social media platforms such as Nextdoor and local Facebook groups.

\paragraph{Challenge of Online Discussions}

Some pro-casino citizens perceived implicit threats from opposition groups on social media, leading to unpleasant interactions. On Nextdoor, opposition groups expressed concerns about introducing a casino locally, while pro-casino citizens emphasized the proposal's potential community benefits. During these discussions, opposition members promoted an online petition on their Facebook page and invited Nextdoor users to join and sign it. However, when a pro-casino participant posted supportive comments about the casino in this Facebook group, opposition members privately messaged them and asked them to stop commenting.
Opposition groups thus appeared to enforce implicit norms against pro-casino expression, showing little tolerance for differing views. Moreover, pro-casino citizens hesitated to express their views publicly on non-anonymous platforms such as Nextdoor, fearing negative labels. For example, P8 noted,

\begin{quote}
\textit{It's like everybody's afraid to speak their mind because they're afraid that they're going to be labeled as a bad person if you don't agree with what's moral. They're afraid to say something publicly. Because we don't want to fight with anybody. - P8 (favor)}
\end{quote}

Many participants viewed existing social media platforms as ineffective venues for productive community discussions. Some believed that platforms like Nextdoor amplified negativity and extreme positions and reflected only vocal minorities. Consequently, some pro-casino residents felt that broader community acceptance was being overshadowed by online opposition, a dynamic that motivated P9 (favor) to offer a more balanced counter-narrative: \emph{``Oh, let's be reasonable, in my opinion, I think the majority of citizens are accepting of a casino and that the app allows a few people to express individual concerns.''}

Another issue was the absence of neutral attitudes during local discussions. Some citizens felt that extreme positions polarized discussions and intensified negative emotions. P10 (neutral) described this phenomenon as typical of contemporary American discourse:

\begin{quote}\emph{Everything is one side or the other. There the people who cannot stand in casinos and hate them and want them nowhere. And then there's the people that say, Oh, I love going to casinos. Unfortunately, I think most discussions today in this country are Me versus You. It's challenging. - P10 (neutral) }\end{quote}

\section{Discussion}

In this section, we discuss conceptual contributions related to sense of misinformation, including strategies for recognizing this distinct phenomenon and repairing it. We then discuss design approaches that could help mitigate SoM.

\subsection{Sense of Misinformation Is Not Misinformation per se but It Is Harmful}

We introduce the concept of \textit{sense of misinformation} (SoM) as the mistaken attribution of misinformation to someone's language or behavior when there is no actual misinformation. Distinguishing misinformation from SoM can be difficult because they can have similar interactional consequences, including undermining community trust, coordination, and collaboration. But they are conceptually distinct. Misinformation involves false information \cite{mcclure2020misinformation}, whereas SoM involves the mistaken perception of false information where none exists. This distinction is important because the two call for different forms of repair. Misinformation requires correcting false information \cite{vraga_correction_2020, cook2015misinformation}, whereas SoM requires recognizing and repairing the misattribution itself.

In our case study, SoM did not emerge suddenly. Rather, it developed over time through repeated breakdowns in coordination and communication, mounting frustration, and deteriorating civility among state government, local government officials, and community members. This is important to note because it suggests that SoM was not the result of an intentional act of deception or a deliberate effort to spread falsehoods \cite{starbird_disinformation_2019}. Instead, it arose incidentally within a difficult civic process, as ongoing disagreement, frustration, and confusion led community members to misattribute others' information as misinformation. It was not merely a momentary misunderstanding between parties, but a socio-relational problem that developed within a fraught local civic process.


SoM carried relational and emotional costs for the community. Our case study suggests that pervasive SoM undermined trust, strained interpersonal relationships, and made collaboration more difficult. Such erosion can be especially consequential in local communities, where collaborative conflict resolution is important for maintaining social capital \cite{coleman_community_1957}. Civic engagement is vital for fostering mutual understanding \cite{resnick_bursting_2013, paffenholz_civil_2006}, but when SoM becomes pervasive, it can weaken the interpersonal ties necessary for sustained social cohesion \cite{carroll_wild_2013, kwon_localized_2020}. Because democratic participation depends on community-level deliberation, trust and civic engagement, breakdowns in local civic processes can also undermine democracy at broader scales \cite{mccord_beyond_2023, sistrunk2024redistrict, nelimarkka2019review}.

Our case study makes two conceptual contributions to the GROUP and HCI communities: (1) it introduces \textit{sense of misinformation} as a distinct interactional phenomenon from misinformation itself, and (2) it shows why this distinction is important for HCI. 
Even when there is no misinformation, mistaken perception of misinformation can still undermine trust, collaboration, and community dialogue. Repairing SoM requires mechanisms and channels that help community members reconcile misattributions of misinformation, and reflect critically on their own judgments.
The HCI community can help raise awareness that one's strong perception of misinformation may, in fact, be mistaken. This important step could help people recognize that such mis-perceptions can occur in severe community conflict. This is exemplified by one township official who reflected on their own remarks and corrected themselves (e.g., \emph{``maybe it’s not misinformation... (P12)''}). These examples suggest that awareness, self-reflection, and critical thinking may help repair SoM, for example through establishing educational initiatives or collaborating with public libraries to promote information literacy efforts.

Finally, our work also encourages the HCI community to explore whether or how sense of misinformation arises in the context of local community conflicts that involve deeply held, divergent community values and beliefs. Our case study shows that once information is perceived as misinformation, it can pose a significant challenge in distinguishing truth from misinformation. This misattribution of misinformation can complicate conflicts, making communication nearly impossible and leading people to mistrust each other more than they should. In our case study, a pervasive SoM suggests a deterioration of these norms, leading to environments of animosity. In such divisive community challenges, future HCI research needs to investigate how to support the re-establishment of community norms, particularly regarding what constitutes appropriate civil behavior.

\subsection{Implications for Design and Practice}
In this section, we discuss how key factors (i.e., miscoordination governmence, miscommunication between local government and citizens and conflict and the breakdown of civic discourse) contributed to sense of misinformation and suggest design approaches that can support communities in mitigating SoM.
\renewcommand{\arraystretch}{1}

\newcommand{\cellsep}{\par\vspace{0.5em}}

\begin{table}[h!]
\centering
\begin{tabular}{p{4cm} p{4.3cm} p{5cm}}
\hline
\textbf{Stakeholder Relationship} & \textbf{Key Findings} & \textbf{Implications} \\
\hline

\textbf{5.2.1. Miscoordination Governance (State vs. Local Gov)} &
The state government imposed a tight six-week ``opt-out'' deadline on the township. This forced a rushed internal decision with minimal public input [4.2.1] &
1.\;Enable intergovernmental dialogue and procedural flexibility: Create shared intergovernmental communication channels to move beyond one-way information dissemination and allow local governments to request deadline extensions for contentious issues \cite{emerson2012integrative, hooghe2010types, gil2016government, ansell2008collaborative, kincaid2017eclipse}\\%
\hline

\textbf{5.2.2. Miscommunication (Local Government vs. Citizens)} &
1.\;The local government's communication infrastructure was fragile (e.g., website updates caused the loss of email subscribers) [4.2.2]%
\cellsep
2.\;Public hearings lacked feedback; citizens felt unheard, citizens became frustration, and local government officials perceived hostility [4.2.2]%
\cellsep
&
1.\;Build trust through ongoing feedback: Implement post-hearing feedback loops (e.g., on the government website) to show that citizens' input is valued \cite{corbett2019towards}%
\cellsep
2.\;Center community storytelling: Integrate online discussion forums into the local government's website to create space for neighborhood storytelling \cite{kim2006community, carroll_neighborhood_2014, coleman_community_1957, disalvo_making_2014}\\
\hline

\textbf{5.2.3. Conflict and the Breakdown of Civic Discourse (Citizen vs. Citizen)} &
1.\; Citizens developed collective arguments and acted collectively [4.1.1]
\cellsep
2.\; Online discussions on Nextdoor and Facebook were hostile, polarized, and marked by a me-versus-you dynamic [4.2.3]%
\cellsep
\cellsep
&
1.\;Visualize public opinion in social context: Represent diverse citizens' views on a 2-D map with demographic identities information to support perspective-taking across disagreement \cite{kim2021starrythoughts}%
\cellsep
2.\;Support inclusive civic discourse: Complement existing online spaces with a community-moderated, neutral public forum that encourages broader participation%
\cellsep
3.\;Encourage prosocial interaction: Use real-time feedback systems to flag uncivil language, prompt self-reflection, and offer paraphrasing suggestions \cite{chang2022thread}\\
\hline

\end{tabular}
\caption{Summary of key findings and corresponding design and practice implications across three stakeholder relationships: (1) state and local government miscoordination, (2) local government and citizen miscommunication, and (3) citizen and citizen conflict and breakdowns in civic discourse}
\label{tab:stakeholder-findings-implications}
\end{table}
\subsubsection{Miscoordination Governance (State VS. Local Government)}

Intergovernmental cooperation is important for the delivery of essential resources and services to the public \cite{meyerhoff_nielsen_digital_2023, peters_challenge_2018}. As discussed in Section \ref{find:miscoordination}, our findings reflected these challenges, particularly in the limited coordination between local and state governments (e.g., the six-week opt-out deadline). Structural and bureaucratic constraints within this multilevel governance system necessitated rushed internal decision-making, which marginalized public input. Because some residents were unaware of these systemic constraints, they interpreted the lack of transparency as local officials deliberately concealing information. This initial misperception fostered deep skepticism.

\textit{Enable intergovernmental dialogue and procedural flexibility:} Effective multilevel governance benefits from interactive dialogue and shared understanding across levels of government \cite{emerson2012integrative, ansell2008collaborative, hooghe2010types}. In contentious contexts, one-way dissemination of information (e.g., the state imposing tight deadlines on local governments) may therefore be insufficient. One possible design direction is dedicated intergovernmental digital platforms that foster mutual awareness of institutional needs and priorities before deadlines are finalized \cite{gil2016government, ansell2008collaborative}. However, new technology alone may not be sufficient when procedural constraints remain rigid. Providing local governments with greater flexibility, such as the ability to negotiate deadline extensions for potentially contentious community issues, is also essential \cite{kincaid2017eclipse}. Such procedural flexibility could help municipalities begin public engagement earlier, better explain institutional constraints, and reduce some citizens' perceptions that local officials are concealing information, thereby mitigating the sense of misinformation that emerged from rushed decision-making.

\subsubsection{Miscommunication between Local Government and Citizens (Local Government VS. Citizens)}

As discussed in Section \ref{find:miscommunication}, the municipality's information-sharing practices were not effectively integrated with its existing communication infrastructure, making it difficult to reach a broad public during the decision-making process. For some citizens, the lack of timely communication was interpreted as evidence that the government was either indifferent or concealing information, leading opposition groups to turn to alternative information practices. Although the government posted factual information on its website, citizens viewed the hearings as performative and noted little genuine interest or feedback on public concerns \cite{fan2022access}. The lack of direct communication and meaningful feedback left anti-casino citizens feeling undervalued, which may escalate conflict and weaken social cohesion \cite{coleman_community_1957}. This aligns with what HCI scholars describe as transactional democracy, in which citizens' agency is limited to one-off participation and they cannot meaningfully shape decision-making processes \cite{mccord_beyond_2023}. Local government officials, meanwhile, sometimes perceived citizens' hostility and heightened emotions as personal, irrelevant, or even as misinformation. These mutual misattributions of misinformation ultimately undermined meaningful democratic discussion within the community.
 
\textit{Build trust through ongoing feedback:}
Active engagement between governments and citizens is essential for rebuilding trust and strengthening civic dialogue \cite{cheng_feedback_2022, mccord_beyond_2023}. However, it is often infeasible for local governments to respond to every individual concern during public hearings. Building on \citet{corbett2019towards}, who conceptualize trust in digital civics as an ongoing sociotechnical process of reducing distance in civic relationships, one potential approach is to implement post-hearing feedback loops on local government websites. These feedback loops could transform a public hearing from a one-time event into an ongoing, bidirectional interaction, potentially supporting the building and retention of trust over time. When residents see their concerns visibly acknowledged and addressed, they may be less likely to respond with animosity, which can help officials interpret citizen input more constructively. Together, these changes can reduce sense of misinformation and support more effective democratic engagement in the community.

\textit{Center community storytelling:} Providing factual information alone is not enough, especially in local communities where meaningful civic engagement often grows out of shared neighborhood stories and attention to people's lived experiences \cite{kim2006community, carroll_neighborhood_2014, coleman_community_1957}. \citet{disalvo_making_2014} argued that matters of concern are important to democracy because democracy is shaped not only by objective facts, but also by how issues are felt and connected to consequences in everyday life. In our case study, the local government provided comprehensive factual information with the altruistic aim of improving the community's understanding of the casino proposal. Yet the website offered no space for engaging with citizens' personal narratives. One practical way to address this is to incorporate even a simple online discussion forum into local government websites. Such a forum could allow both officials and residents to share their experiences, and could help strengthen social capital and community bonds \cite{cong2021collective, lopez2015lend}. Strengthening community bonds through neighborhood storytelling may help community members respect differing values rather than misattribution of misinformation.

\subsubsection{Conflict and the Breakdown of Civic Discourse (Citizen VS. Citizen)}
As discussed in Section \ref{find:conflictbreakdown}, sense of misinformation became pervasive among community members and contributed to an atmosphere of animosity. Rather than interpreting disagreements as legitimate differences in values or beliefs, both sides often perceived each other's information as misinformation. This dynamic shaped opposition citizens' information practices, including their website and offline petitioning \cite{kaviani2022bridging}, which some fellow citizens perceived as false (i.e., SoM) and threatening. These tensions also extended to online community platforms that are often intended to support community building and collective efficacy \cite{virtanen2008supporting, choksi2024under, carroll_collective_2005}. Opposition citizens' local Facebook groups strengthened internal solidarity while leaving pro-casino citizens feeling excluded, and discussion on Nextdoor \cite{NextdoorAbout} was often marked by tension rather than respectful communication. As a result, community stakeholders faced significant challenges in discussing the casino proposal.

\citet{coleman_community_1957} suggests that community conflict can enter a downward spiral, leading to extreme polarization and weakened social relations that may become irreversible. Building on \citet{coleman_community_1957}'s classic concept of community conflict, our case study showed that sense of misinformation can further complicate such conflicts. These dynamics can also divide communities and make meaningful communication nearly impossible.

\textit{Visualize public opinion in social context:} Currently, there is a lack of online spaces that allow the public to explore diverse perspectives in a structured and respectful manner. Exposure to diverse opinions is important \cite{kriplean_supporting_2012}, but exposure alone may not be sufficient to mitigate polarization or foster genuine respect for opposing viewpoints. 
One potential approach is to visualize wide range of public opinions as individual dots in a two-dimensional space while also providing social context such as demographic information. Prior work suggests that presenting opinions alongside demographic identities can help people explore diverse perspectives, understand the contexts behind community members' views, and empathize with the experiences that shape those views \cite{kim2021starrythoughts}.
Such a design may create opportunities for people to interpret disagreement as a reflecting differing beliefs and values rather than misinformation.


\textit{Support inclusive civic discourse:} Our case study revealed that anti-casino residents formed homogeneous local Facebook groups to strengthen internal solidarity. However, when such forums primarily serve as only venues for discussion, they can reinforce confirmation bias \cite{mercier_confirmation_2022, mercier2022confirmation}, intensify cognitive dissonance \cite{oconnor_analysis_2017, festinger1962cognitive}, and limit engagement with opposing views. This may ultimately weaken communication across the broader community. Without dismissing the spaces that citizens create for themselves, local governments could complement them with a well-moderated, community-wide public forum. Building on prior HCI research \cite{gao_burst_2018,kriplean_supporting_2012}, the purpose of such a forum is to offer an accessible and neutral space where all residents can engage respectfully across differences, thereby reducing the tendency to interpret disagreement as misinformation.

\textit{Encourage prosocial interaction:} Respect and tolerance for differences are essential for sustaining civic dialogue and protecting civil liberties \cite{walsh2007democratic, mccoy2002deliberative}. To promote respectful community norms online, local governments might consider mechanisms that gently prompt users before they post in online discussions, helping prevent tense situations from escalating. For example, real-time feedback systems could flag potentially tense or uncivil phrasing and suggest revisions that encourage empathy and prosocial interaction \cite{chang2022thread}. Although such tools may be economically or bureaucratically difficult for municipalities to develop, lighter-weight interventions, such as community moderation protocols, may still help citizens recognize and manage the risks of escalation. By reducing animosity, prosocial interactions may help lower collective mistrust and reduce sense of misinformation.

\subsection{Limitations and Future Work}

This study has several limitations. First, our participants were recruited from a single local community and represented only a small group of its residents. We therefore avoid making strong generalizations to other populations. Future work should investigate additional local communities to explore our results. Second, we focused on a single context; a casino development proposal. Future HCI research could explore how or whether the concept of sense of misinformation applies in other community settings. Finally, we did not collect demographic information such as age, education, or occupation. These factors may have shaped differences in stakeholders' perspectives and information practices. Future work should investigate how demographic factors influence information practices and experiences of sense of misinformation in local community contexts.


\section{Conclusion}
Through a case study of a local casino development proposal, we introduce the concept of \textit{sense of misinformation} as a phenomenon distinct from actual misinformation. We identify several factors that contributed to this sense of misinformation, including miscoordination between state and local governments, miscommunication between local government and citizens, and conflict and the breakdown of civic discourse. Our work contributes to the misinformation literature by identifying how the misattribution of misinformation can emerge over time through repeated frustration, confusion, and deteriorating civility. This work also extends prior HCI research on misinformation, which has often focused on global level misinformation and cases involving actual misinformation.

\section{Acknowledgment}
All authors sincerely appreciate the local government officials and citizens for their generous time and support. We are also deeply grateful to the reviewers for their thoughtful and insightful feedback. We extend special thanks to Jessica Perkins for immense help in improving an early draft of the paper, and Kim extends heartfelt gratitude to Jessica Perkins for unwavering encouragement and belief throughout the entire project journey, particularly during periods of doubt. The authors also thank Jane Im for invaluable feedback during the revision phase and for steadfast support. The authors also thank Ginny Wang for insightful feedback that strengthened the wording of the research questions and for emotional support, and thank Hongyi Dong for helpful feedback during the writing process.

\bibliographystyle{ACM-Reference-Format}
\bibliography{sense_misinfo, jiyoon_reference}










\end{document}